\documentclass[a4paper]{jpconf}
\usepackage{graphicx}
\begin{document}
\title{Nuclear incompressibility from spherical and deformed nuclei}

\author{
Danilo Gambacurta$^a$, 
Gianluca Col\`o$^{b,c}$, Alessandro Pastore$^d$}

\address{
$^a$Extreme Light Infrastructure - Nuclear Physics (ELI-NP), Horia Hulubei National Institute for Physics and Nuclear Engineering,
30 Reactorului Street, RO-077125 Magurele, Jud. Ilfov, Romania\\
$^b$Dipartimento di Fisica ``Aldo Pontremoli'', Universit\`a degli Studi di Milano, 20133 Milano, Italy\\
$^c$INFN, Sezione di Milano, 20133 Milano, Italy\\
$^d$Department of Physics, University of York, Heslington, York, YO10 5DD, UK
}

\ead{Gianluca.Colo@mi.infn.it}

\begin{abstract}
We present an analysis based on the deformed Quasi Particle Random Phase Approximation, on top of a deformed Hartree-Fock-Bogoliubov description of the ground state, aimed at studying the isoscalar monopole and quadrupole response in a deformed
nucleus. This analysis is motivated by the need of understanding the coupling between the two modes and 
how it might affect the extraction of the nuclear incompressibility from the monopole distribution. After discussing this motivation, we present the main ingredients of our theoretical framework, and we show some results
obtained with the SLy4 and SkM$^{*}$ interactions for the  nucleus ${}^{24}$Mg. 
\end{abstract}

\section{Introduction}
The nuclear Equation of State (EoS) plays a fundamental role in the description and understanding of several properties of atomic nuclei and astrophysical compact objects~\cite{RevModPhys.89.015007}. However, its knowledge is, from a quantitative point of view, far from  being sufficiently enough well determined. One of the key quantities needed to better constrain the EoS is the nuclear matter incompressibility, \emph{e.g.} the curvature of the symmetric nuclear matter EoS, which is closely related to the compressional modes of atomic nuclei~\cite{bla80}.  

The emergence of collective excitations is a common  feature of many-body systems. In atomic nuclei, the most collective excitations are the Giant Resonances (GRs), described as a coherent motion to which many nucleons participate \cite{Harakeh_book}. The Isoscalar Giant Monopole Resonance (ISGMR), known also as the breathing mode, is a typical example of collective excitation whose centroid energy is related to the nuclear incompressibility. The properties of GRs are also considerably affected by the shape of the nucleus and in particular by its deformation. A very well known example is  the Isovector Giant Dipole Resonances, whose strength distribution is split in two peaks, corresponding to a vibration along or perpendicular to the symmetry axis. The distance in energy between these two peaks is related to the corresponding deformation of the nucleus~\cite{mar16}.

Deformation also affects the ISGMR and we aim at dicussing such effects in view of existing and future experimental results on deformed nuclei.
The available measurements of the ISGMR done in closed-shell nuclei 
(mainly $^{208}$Pb) tend to point toward a value of nuclear compressibility of  $K_{\infty}\approx$ 240 $\pm$ 20 MeV~\cite{gar18}, while other measurements done in open-shell nuclei  point to lower values of the nuclear matter incompressibility and the effect of deformation has still to be investigated.

In the case of the ISGMR, deformation produces a coupling of the  K = 0 
component of the Isoscalar Giant Quadrupole Resonance (ISGQR) with the 
ISGMR itself, K being the projection of the total angular momentum on the symmetry axis~\cite{rag05}. How this coupling might affect the extraction of the nuclear incompressibility is still to be investigated and understood. 
In this work, we show some preliminary results in this direction.

\section{Methodology and Results} 
\begin{figure}[h]
\begin{minipage}{18pc}
\includegraphics[width=20pc]{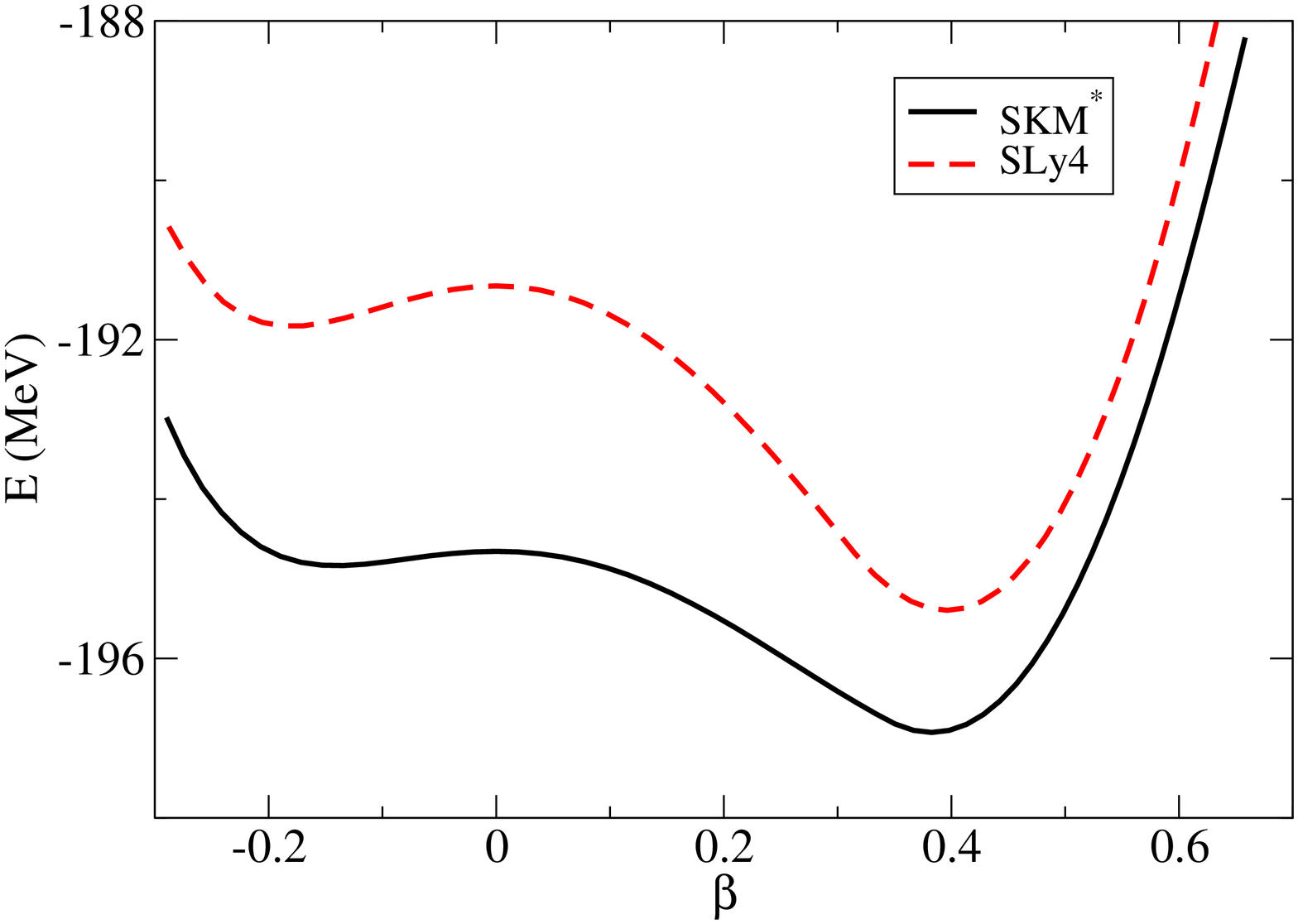}
\caption{\label{Fig_PES}HFB potential energy surface, obtained with the SkM$^{*}$ and SLy4 interactions, as a function of the axial deformation parameter $\beta$, defined in Eq. (\ref{beta tot}), for  ${}^{24}$Mg.}
\end{minipage}\hspace{2pc}%
\begin{minipage}{18pc}
\includegraphics[width=20pc]{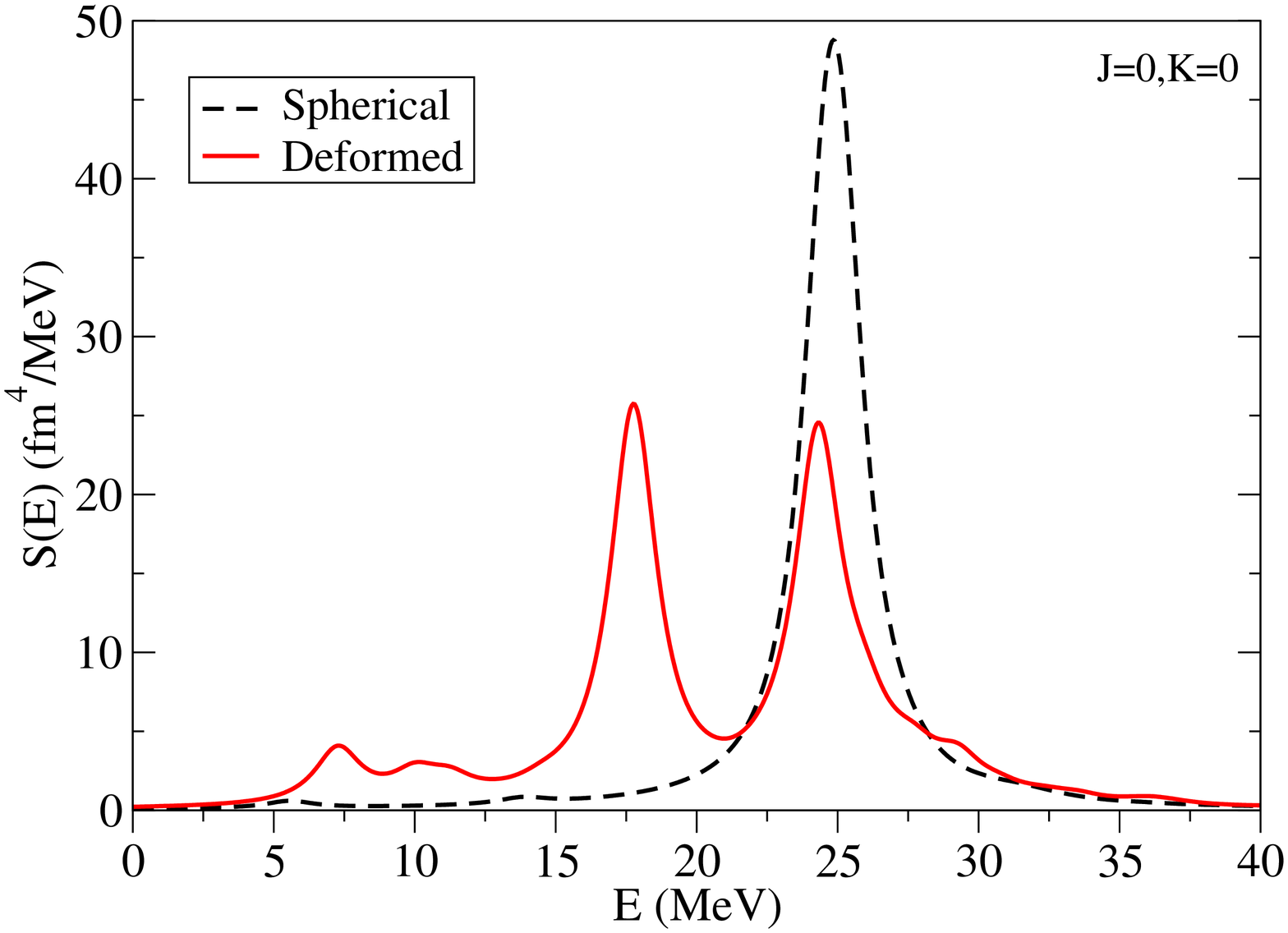}
\caption{\label{Fig_Str_SLY4_SphvsDef} Monopole  strength functions obtained for the SLy4 force. Red solid (black dashed) lines correspond to the deformed (spherical) solution.}
\end{minipage} 
\end{figure}

We briefly  summarise the main features of our calculations, and for more details we refer to  Ref. \cite{Losa}. The first step in our methodology consists in solving the Skyrme- Hartree-Fock-Bogoliubov (HFB) equations  for an axially deformed system in a finite  deformed harmonic oscillator basis by using the HFBTHO code \cite{HFBTHO}. A density-dependent pairing delta interaction, that is, 
\begin{equation}\label{pairing int}
V_{pair} \left( {\textit{\textbf{r}}, \textit{\textbf{r}}'} \right)  = V_0\left[ {1  +
\gamma \frac{ \rho \left( \textit{\textbf{r}} \right)}{\rho_c} } \right]\delta \left( {\textit{\textbf{r}} -
\textit{\textbf{r}}'} \right)\;,
\end{equation}
is used in the particle-particle channel. 
The values of the parameters are $V_{pair}=-280$ MeV fm$^3$, $\rho_c=0.16 fm^{-3} $ and $\gamma=0.5$.
In the particle-hole channel we use a standard Skyrme functional~\cite{ben03}.

The total energy is minimized  under the constraint of a fixed quadrupole moment to which we associate the $\beta$ parameter  defined as
\begin{equation}\label{beta tot}
\beta  = \sqrt {\frac{\pi } {5}} \frac{{ < \hat Q > _n  +  < \hat Q > _p }} {{\left\langle {r^2 } \right\rangle _n  +
\left\langle {r^2 } \right\rangle _p }}, 
\end{equation}
\noindent  where $< \!\! \hat Q \!\! >_q$ is the average value of the quadrupole--moment operator  for protons ($p$) and neutrons ($n$). 
This step allows to determine the $\beta$ value for which the minimal energy of the system is obtained. The  potential energy curves for the SkM$^{*}$ \cite{SKMS} and SLy4 \cite{SLY4} interactions are shown in Fig. \ref{Fig_PES} as a function of $\beta$. In both cases we see a minimum corresponding to the prolate deformation $\beta\approx 0.4$.  For this particular choice of interactions, we observe that  the 
solution for the ground state is not superfluid.

The next step consists in calculating the nuclear response  by using the  Quasi Particle Random Phase Approximation (QRPA) approach~\cite{suh07}. The QRPA calculation is carried out on top of the HFB solution corresponding to the absolute minimum of the potential energy.
The QRPA excited states $\left| \lambda  \right\rangle$ are described in terms
of the excitation operators
\begin{equation}\label{Qdaga}
Q_\lambda ^\dag   = \sum\limits_{K < K'} {\left( {X_{KK'}^\lambda  \alpha _K^\dag  \alpha _{K'}^\dag   -
		Y_{KK'}^\lambda  \alpha _{K'} \alpha _K } \right)}\;,
\end{equation}

\noindent normalized as 
\begin{equation}\label{Qdaga}
 \sum\limits_{K < K'} \left( (X_{KK'}^{\lambda })^2 - (Y_{KK'}^{\lambda})^2 \right)=1\;,
\end{equation}

\noindent where ${\alpha _K^\dag  }$, ${\alpha _K }$ are the quasiparticle creation and annihilation operators,
respectively, and the pairs are such as $\Omega  = \Omega _K  + \Omega _{K'}$ and $\pi  = \pi _K \cdot \pi _{K'} $. Here, $\Omega$ is the projection of the angular momentum on the symmetry axis $z$ and $\pi$ is the parity.
The $X$ and $Y$ amplitudes together with the energies of the excited states are obtained by solving the QRPA eigenvalue problem  in the matrix form 
\begin{equation}\label{eq_srpa}
\nonumber
\left(\begin{array}{cc}
  A & B \\
  -B^{*} & -A^{*} \\
\end{array}\right)
\left(%
\begin{array}{c}
  X^{\lambda} \\
  Y^{\lambda} \\
\end{array}%
\right)=\omega_{\lambda}
\left(%
\begin{array}{c}
  X^{\lambda} \\
  Y^{\lambda} \\
\end{array}%
\right).
\end{equation}
\begin{figure}[h]
\begin{minipage}{14pc}
\includegraphics[width=20pc]{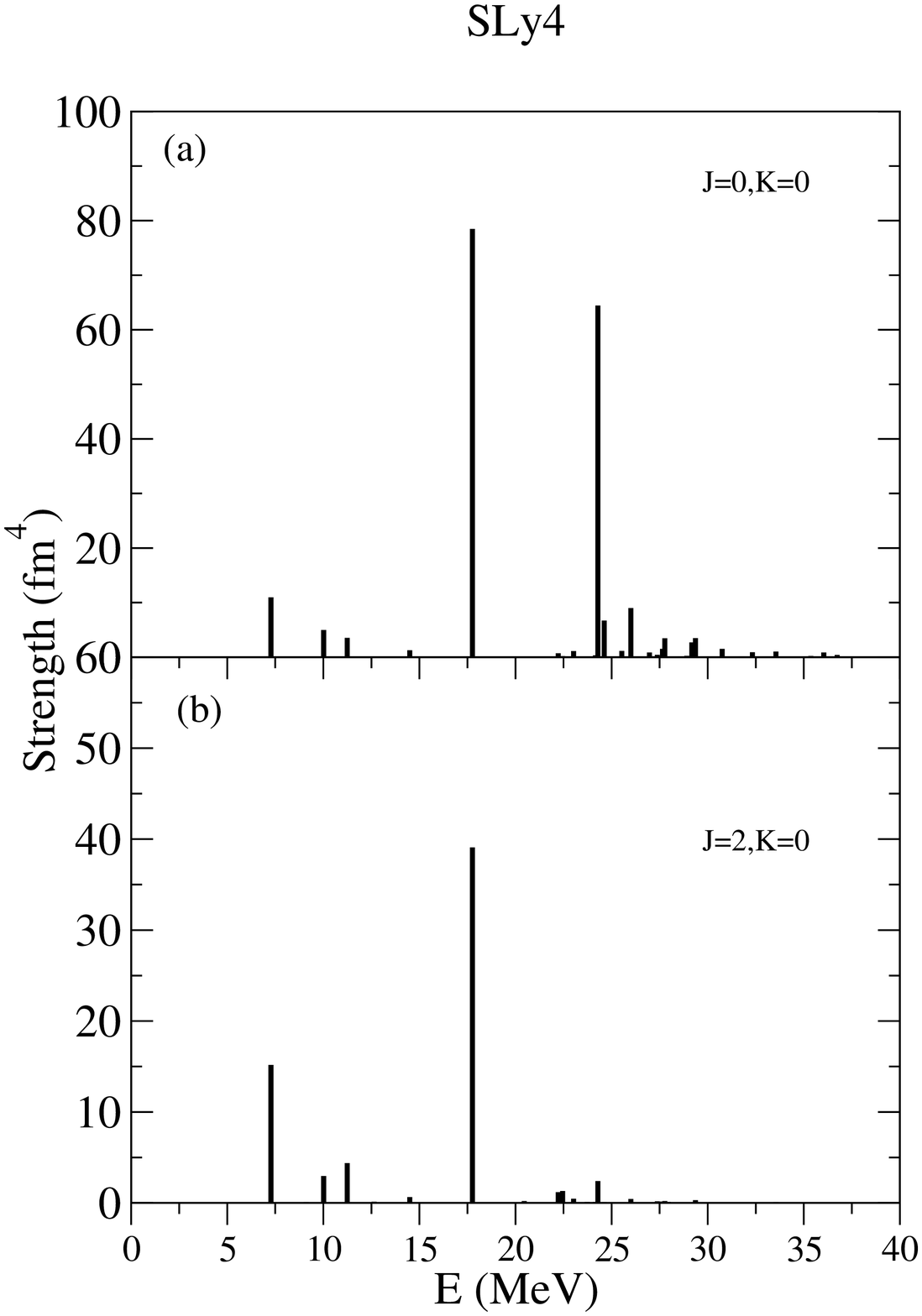}
\caption{\label{Fig_Str_SLY4} Monopole (upper panel) and quadrupole (lower panel) strength distribution obtained for the SLy4 interaction.}
\end{minipage}\hspace{4pc}%
\begin{minipage}{14pc}
\includegraphics[width=20pc]{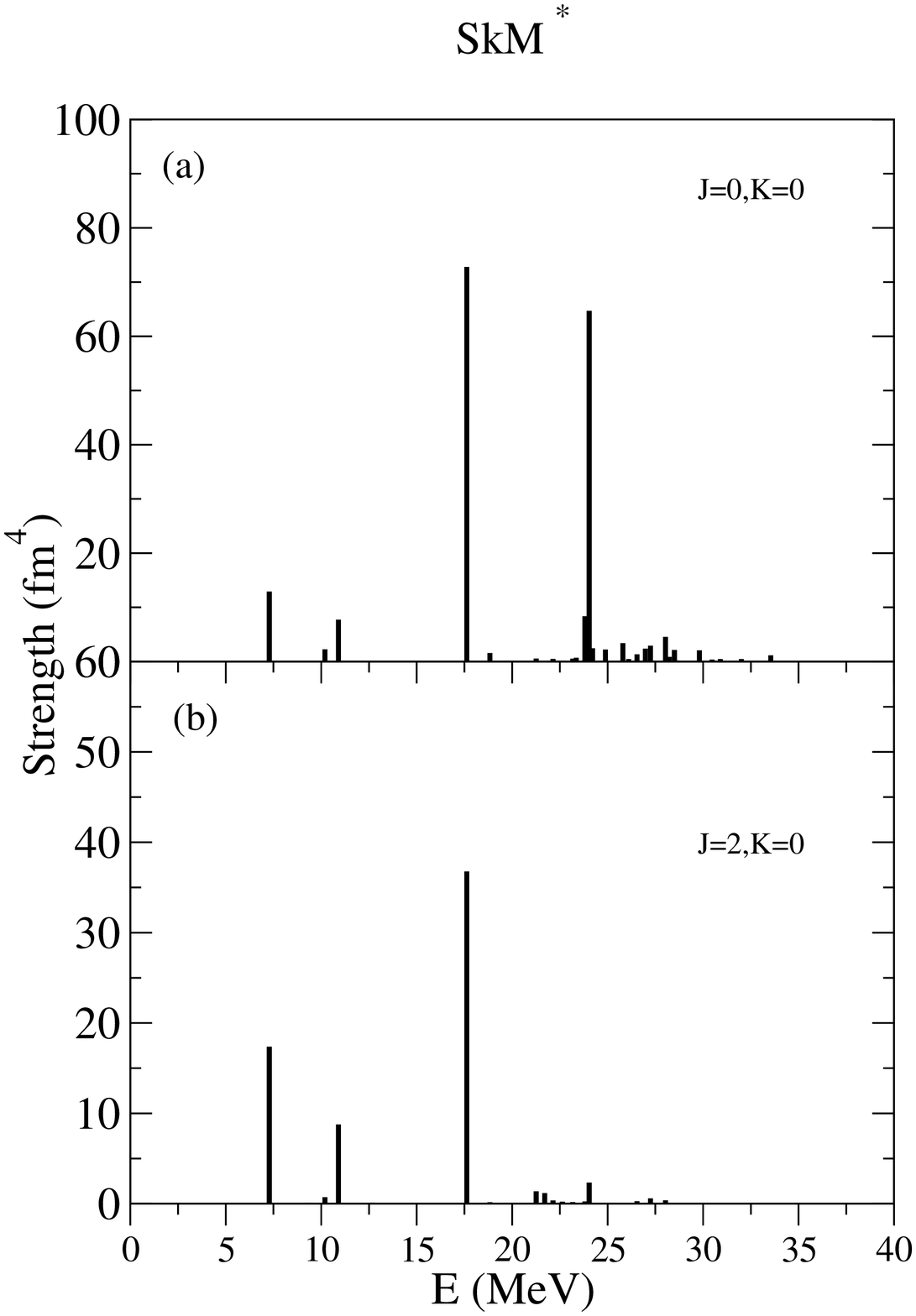}
\caption{\label{Fig_Str_SKMS} Monopole (upper panel) and quadrupole (lower panel) strength distribution obtained for the SkM$^{*}$ interaction. 
}
\end{minipage}
\end{figure}

We will focus here on the   isoscalar quadrupole and monopole  strengths whose transition operators are defined as
\begin{equation}\label{F quadrupolo IS}
\hat F_{J=2,K}^{IS}  = \frac{{eZ}} {A}\sum\limits_{i = 1}^A {r_i^2 Y_{2 K } \left( {\hat r_i } \right),~~~}
\hat F_{J=0,K=0}^{IS}  = \frac{{eZ}} {A}\sum\limits_{i = 1}^A {r_i^2 }.
\end{equation}

In Fig. \ref{Fig_Str_SLY4_SphvsDef} we compare the QRPA results for the SLy4 force, obtained at the deformed minimum $\beta\approx 0.4$ and at the spherical limit $\beta=0$. For sake of comparison, we plot  the folded monopole strength functions 
\begin{equation}\label{lorenziana}
S  \left( E \right) = \sum_{\lambda}  {  {\frac{{\Gamma /2}} {\pi }} } \frac{{\left|
{\left\langle \lambda  \right|\hat F \left| 0 \right\rangle } \right|^2 }} {{\left( {E - E_\lambda  }
\right)^2  + \Gamma ^2 /4}},
\end{equation}
where $\Gamma=2$ MeV has been used as a smoothing parameter 
 The effect of the deformation is clearly seen, as it leads to a splitting of the main peak obtained in the spherical solution, and to the two peaks observed in the deformed case.  

For the sake of further analysis, we plot 
in Figs.~\ref{Fig_Str_SLY4} and \ref{Fig_Str_SKMS}
the unfolded QRPA strength distributions corresponding to isoscalar monopole and quadrupole  operators, in the upper and lower panels, respectively. For both interactions, the QRPA calculations are performed at the deformed minimum shown in Fig. \ref{Fig_PES}. The monopole strength shows, in both cases, two main peaks  located at 17.76 and 24.29 MeV for SLy4 and 17.62 and 24.03 MeV for SkM$^{*}$. The lower state is also excited by the quadropole operator as it can be seen in the lower panels of the two figures. The Energy Weighted Sum Rules are preserved within less than 1\%~\cite{lip89}.

\begin{figure}[h]
\begin{minipage}{14pc}
\includegraphics[width=15pc]{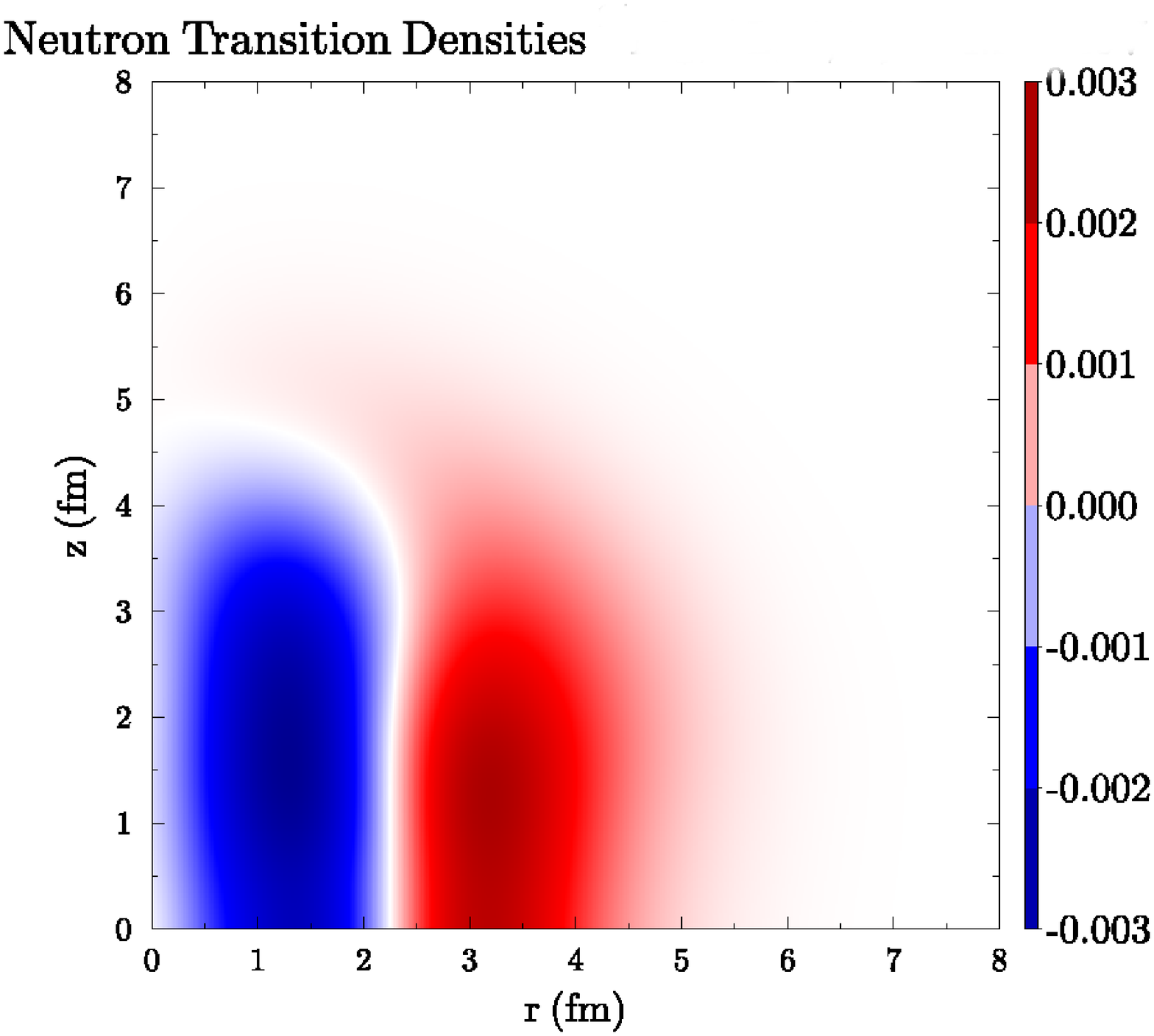}
\caption{\label{Fig_NeutronTD_GMR} Neutron transition density $\delta\rho_\lambda(r,z)$ for the excited state located at 24.29 MeV (see Fig. \ref{Fig_Str_SLY4}).}
\end{minipage}\hspace{4pc}%
\begin{minipage}{14pc}
\includegraphics[width=15pc]{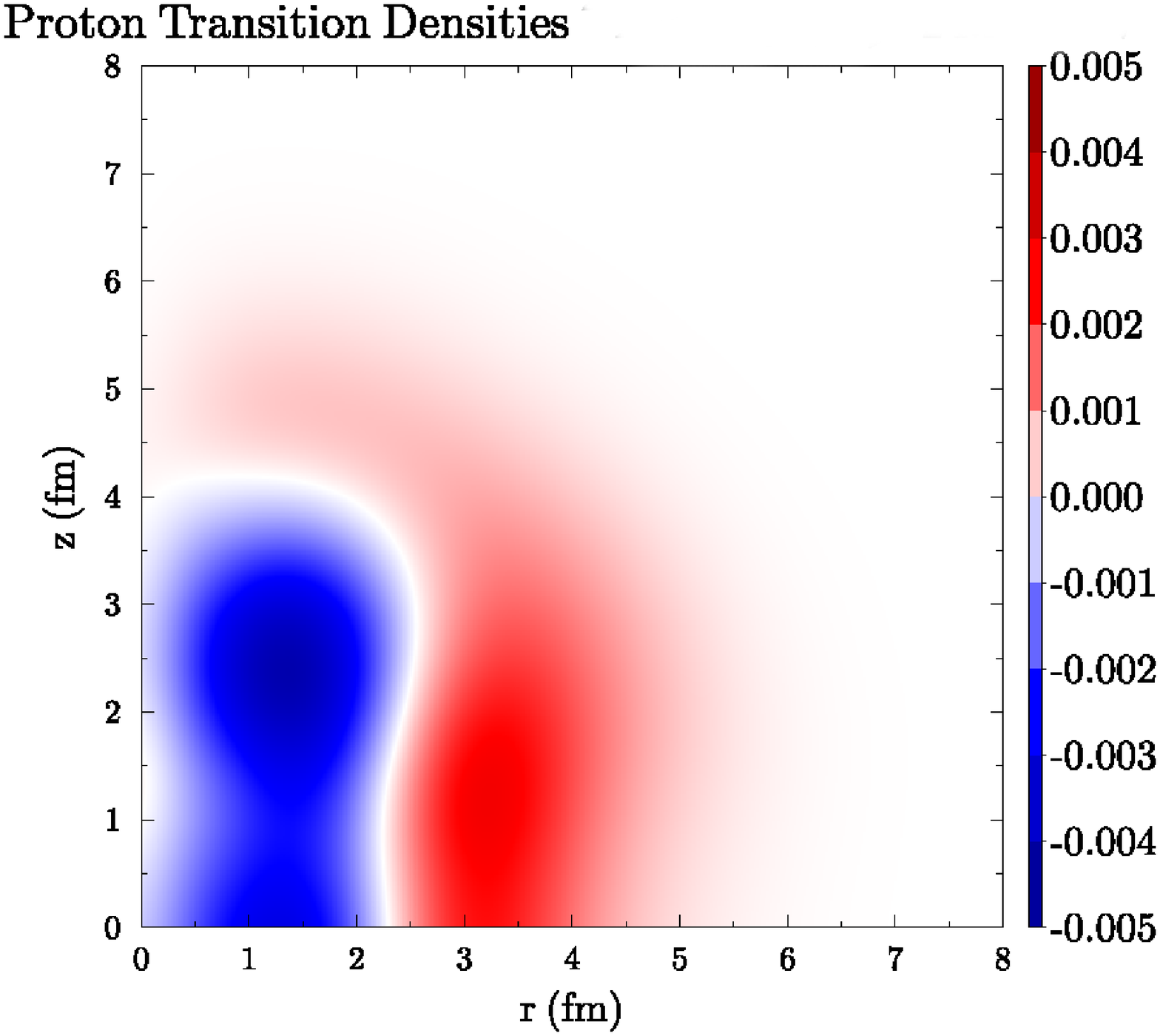}
\caption{\label{Fig_ProtonTD_GMR} Proton transition density $\delta\rho_\lambda(r,z)$ for the excited state located at 24.29 MeV (see Fig. \ref{Fig_Str_SLY4}).}
\end{minipage} 
\end{figure}
In order to have a deeper insight on the nature of these states, we analyze below the transition densities
\begin{equation}\label{TD}
 \rho_\lambda( \vec{r})=\delta\rho_\lambda(r,z)e^{-i K \phi}\;,
\end{equation}
where $K$ is the total angular momentum projection of the excited state and $r,z$ and $\phi$ are the cylindrical coordinates.

\begin{figure}[h]
\begin{minipage}{14pc}
\includegraphics[width=15pc]{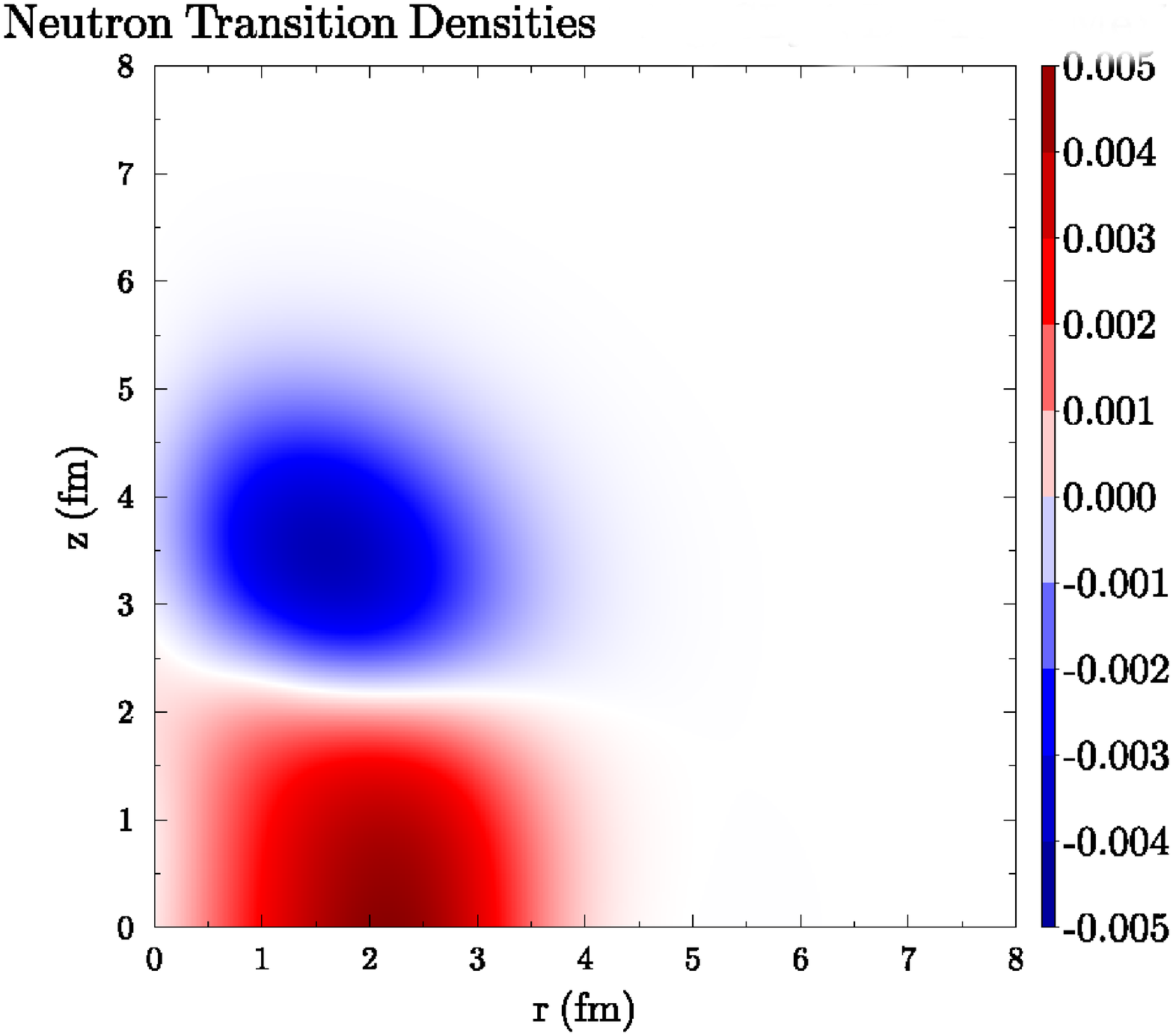}
\caption{\label{Fig_NeutronTD_GQR} Neutron transition density $\delta\rho_\lambda(r,z)$ for the excited state located at 17.76 MeV (see Fig. \ref{Fig_Str_SLY4}).}
\end{minipage}\hspace{4pc}%
\begin{minipage}{14pc}
\includegraphics[width=15pc]{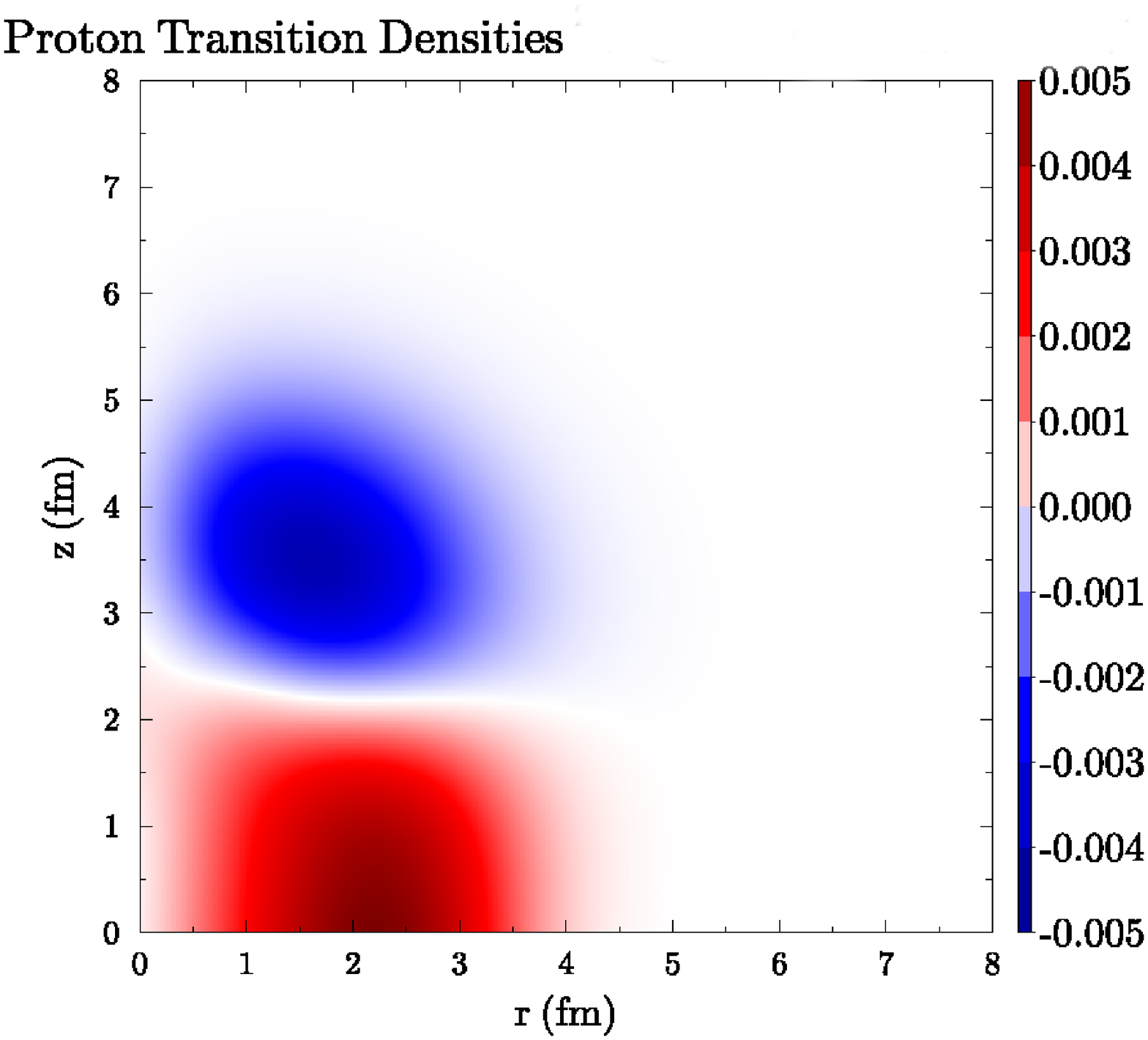}
\caption{\label{Fig_ProtonTD_GQR}Proton transition density $\delta\rho_\lambda(r,z)$ for the excited state located at 17.76 MeV (see Fig. \ref{Fig_Str_SLY4}).}
\end{minipage} 
\end{figure}

In Figs. \ref{Fig_NeutronTD_GQR} and \ref{Fig_ProtonTD_GQR}, we display the neutron and proton transition densities obtained in the SLy4 case for the state at 17.76 MeV, while in Figs. \ref{Fig_NeutronTD_GMR} and \ref{Fig_ProtonTD_GMR} we show the neutron and proton transition densities for the state located at  24.29 MeV.

\begin{figure}[h]
\begin{minipage}{14pc}
\includegraphics[width=15pc]{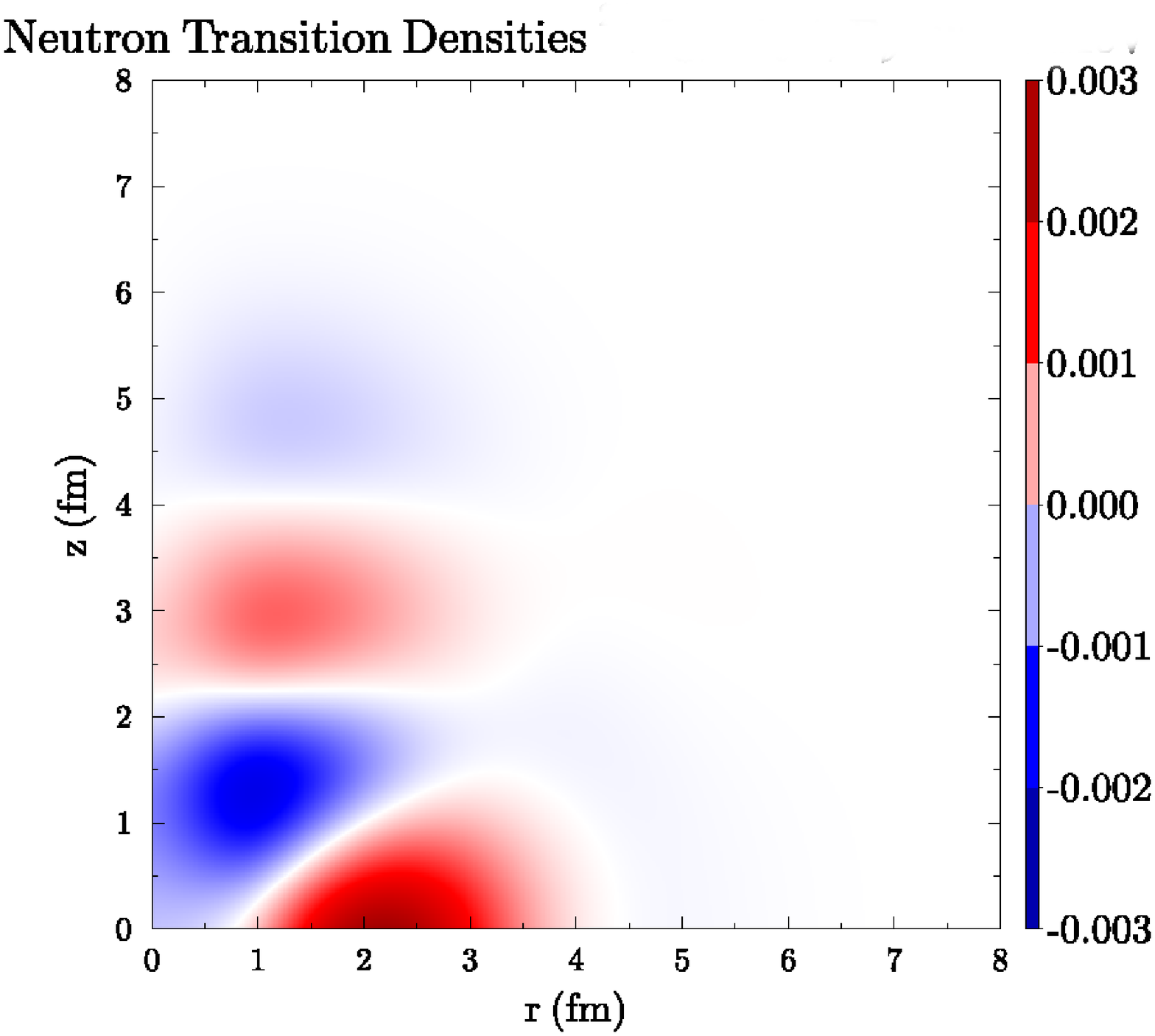}
\caption{\label{Fig_NeutronTD_NC} Neutron transition density $\delta\rho_\lambda(r,z)$ for the (non collective) excited state located at 25.46 MeV (see Fig. \ref{Fig_Str_SLY4}).}
\end{minipage}\hspace{4pc}%
\begin{minipage}{14pc}
\includegraphics[width=15pc]{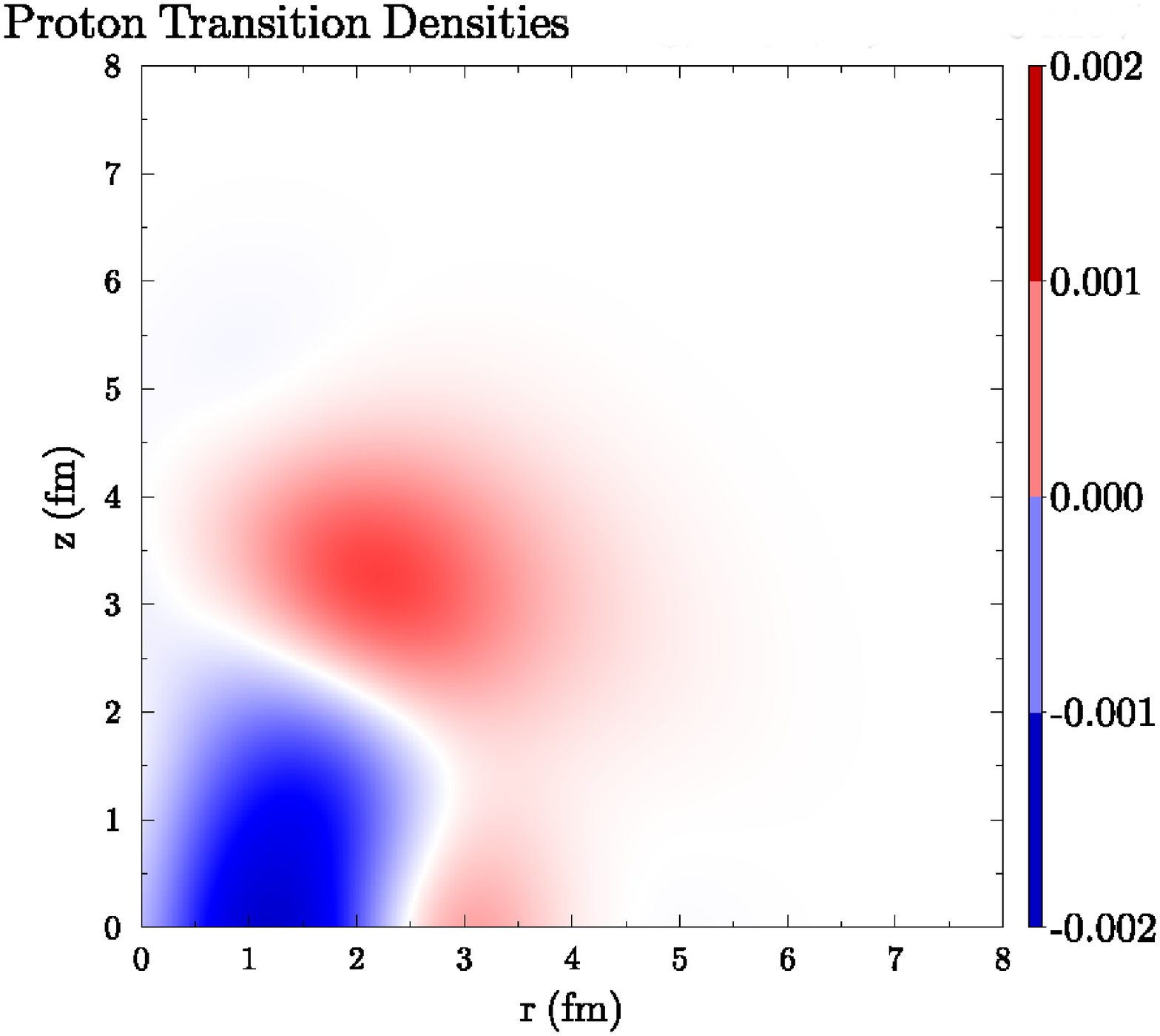}
\caption{\label{Fig_ProtonTD_NC} Proton transition density $\delta\rho_\lambda(r,z)$ for the (non collective) excited state located at 25.46 MeV (see Fig. \ref{Fig_Str_SLY4}).}
\end{minipage} 
\end{figure}

For both states, we observe the expected behaviour for isoscalar excitations, \emph{i.e.} neutrons and protons oscillating in phase. In the case of the monopole excitation (Figs \ref{Fig_NeutronTD_GMR} and \ref{Fig_ProtonTD_GMR}) the oscillation is along the radial axis, while in  the the quadrupole case (Figs \ref{Fig_NeutronTD_GQR} and \ref{Fig_ProtonTD_GQR})
the nucleons oscillate along the axial $z$ axis. For comparison, we plot also, in Figs. \ref{Fig_NeutronTD_NC} and \ref{Fig_ProtonTD_NC}, the transition densities of a non collective state ($E\approx 25.46$ MeV) that clearly shows a different pattern, characterized by several nodes. 
Similar results have been obtained using the SkM$^*$ interaction.

\section*{Conclusions and Outlook}
We have presented some results, based on a deformed QRPA approach, for the monopole and quadrupole strength distributions in the deformed nucleus ${}^{24}$Mg. 
Our study is motivated by the need 
of understanding the compressional properties of deformed nuclei. To this aim, we 
wish to disentangle the 
compressional properties 
associated with the ISGMR, from 
its coupling with the $K=0$ component of the ISGQR. 

Our 
results show clearly the splitting of the monopole strength due to deformation. 
Moreover, 
the transition densities of the collective states have been shown to be
very instrumental to indicate the character of states (compressional or not).
In the near future, 
we plan to implement a projection on the total angular momentum of the QRPA phonons. 
The projection would in fact allow for 
a clear decoupling the two modes, providing a cleaner way to extract the monopole 
peak energy. Work in this direction is on going. 

\ack

This work has been partially supported by STFC Grant No. ST/P003885/1.
Funding from the European Union's Horizon 2020 research and innovation programme under grant agreement No 654002 is also acknowledged.

\section*{References}
\bibliographystyle{iopart-num}

\bibliography{biblio}
%
%
%
%
%
%

\end{document}